%% file: RR-8644.tex
\documentclass[twoside]{article}
\usepackage[a4paper]{geometry}
\usepackage[latin1]{inputenc}
\usepackage[T1]{fontenc} 
\usepackage{RR}
\usepackage{amsmath}
\usepackage{bbm}
\usepackage{RR}
\usepackage{amssymb}
\usepackage{graphicx}
\usepackage{listings}
\usepackage{multirow}
\usepackage{algorithm}
\usepackage{algorithmic}
\usepackage{url}
\usepackage{hyperref}
\hypersetup{linktocpage}
\usepackage{verbatim}
\usepackage{tikz}
\usetikzlibrary{arrows}
\usepackage{longtable}
\usepackage{caption}
\usepackage{color}
\usepackage{xcolor}
\usepackage{colortbl}
\usepackage{booktabs}
\usepackage{parcolumns}
\usepackage{amsfonts}

\definecolor{lincolngreen}{rgb}{0.11, 0.35, 0.02}
\definecolor{limegreen}{rgb}{0.2, 0.8, 0.2}
\definecolor{napiergreen}{rgb}{0.16, 0.5, 0.0}
\definecolor{beaver}{rgb}{0.62, 0.51, 0.44}
\definecolor{battleshipgrey}{rgb}{0.52, 0.52, 0.51}
\definecolor{bole}{rgb}{0.47, 0.27, 0.23}

\hypersetup{
  linktocpage,
  colorlinks=true,
  citecolor=red,
  filecolor=blue,
  linkcolor=red,
  urlcolor=blue
}

\newcommand{\figref}{Figure \ref}
\newcommand{\lstref}{Listing \ref}

\newenvironment{definition}[1][Definition]{\begin{trivlist}
\item[\hskip \labelsep {\bfseries #1}]}{\end{trivlist}}

\newcommand{\qed}{\nobreak \ifvmode \relax \else
      \ifdim\lastskip<1.5em \hskip-\lastskip
      \hskip1.5em plus0em minus0.5em \fi \nobreak
      \vrule height0.75em width0.5em depth0.25em\fi}

\RRNo{8644}
\RRdate{October 2014}
\RRauthor{
Van Chan Ngo
  \and
Axel Legay
  \and 
Jean Quilbeuf
}
\authorhead{Ngo et al.}
\RRtitle{Dynamic Verification of SystemC with Statistical Model Checking}
\RRetitle{Dynamic Verification of SystemC with Statistical Model Checking}
\titlehead{Dynamic Verification of SystemC with Statistical Model Checking}
\input{RRresume}
\input{RRabstract}
\input{motcles}
\input{keywords}
\RRprojet{ESTASYS}
\RCRennes 

\begin{document}
\makeRR   

\tableofcontents
\newpage

\include{introduction}

\include{systemc}

\include{trace}

\include{implementation}

\include{conclusion}

\bibliographystyle{abbrv}
\bibliography{RR-8644}  

\end{document}

%% file: RRresume.tex
\RRresume{
La mod\'elisation au niveau transactionnel de SystemC a tr\`es bien r\'eussi \`a d\'ecrire le comportement des syst\`emes embarqu\'es en fournissant de haut niveau des mod\`eles ex\'ecutables, dans laquelle beaucoup d'entre eux ont un comportement probabiliste inh\'erente, \`a savoir, des donn\'ees al\'eatoires, des composants fiables. Il est crucial pour \'evaluer l'analyse quantitative et qualitative de la probabilit\'e' de les propri\'et\'es du syst\`eme. Une telle analyse peut \^etre effectu\'ee \`a l'aide de probabilistic model checking. Cependant, cette m\'ethode est impossible de traiter avec les syst\`emes vastes et complexes et travaille directement avec la mod\'elisation des syst\`emes en SystemC, en raison de l'espace explosion de l'\'etat. Dans cet article, nous d\'emontrons l'utilisation r\'eussie de statistical model checking \`a effectuer une telle analyse pour les syst\`emes mod\'elis\'es dans SystemC. Notre cadre de v\'erification permet aux concepteurs d'exprimer une large gamme de propri\'et\'es utiles qui peuvent \^etre analys\'es.
}

%% file: RRabstract.tex
\RRabstract{
Transaction-level modeling with SystemC has been very successful in describing the behavior of embedded systems by providing high-level executable models, in which many of them have an inherent probabilistic behavior, i.e., random data, unreliable components. It is crucial to evaluate the quantitive and qualitative analysis of the probability of the system properties. Such analysis can be conducted by using probabilistic model checking. However, this method is unfeasible to deal with large and complex systems and works directly with systems modeling in SystemC, due to the state space explosion. In this paper, we demonstrate the successful use of statistical model checking to carry out such analysis for systems modeled in SystemC. Our verification framework allows designers to express a wide range of useful properties that can be analyzed.
}

%% file: motcles.tex
\RRmotcle{Runtime Verification, Probabilistic Assertion, Statistical Model Checking, Program Verification, SystemC}

%% file: keywords.tex
\RRkeyword{Runtime Verification, Probabilistic Assertion, Statistical Model Checking, Program Verification, SystemC}

%% file: introduction.tex
\section{Introduction}
Transaction-level modeling (TLM) with SystemC has been become increasingly prominent in describing the behavior of embedded systems~\cite{cch99}, i.e., System-on-Chips (SoCs). It allows complex electronic components and software control units can be combined into a single model, enabling simulation of the whole system at once. 
In many cases, models include probabilistic and non-deterministic characteristics, i.e, random data, reliability of the system's components. It is crucial to evaluate the quantitive and qualitative analysis of the probability of the system's properties. We consider a safety-critical system (i.e., the control system for air-traffic, automotive, and medical device). The reliability and availability model of the system can be considered as a stochastic process, in which it exhibits probabilistic characteristics. For instance, the reliability and availability model of an embedded control system~\cite{knp07} that contains an input processor connected to groups of sensors, an output processor, connected to groups of actuators, and a main processor, that communicates with the I/O processors through a bus. Suppose that the sensors, actuators, and processors can be failed, in which the I/O processors have transient and permanent faults. When a transient fault occurs in a processor, rebooting the processor repairs the fault. The times to failure and reboot's delay of processors are exponentially distributed. Then, the reliability of the system cam be modeled by a continuous-time Markov chain (CTMC)~\cite{tks82,gaa87} that is a special case of a discrete-state \textit{stochastic process} in which the probability distribution of the next state depends only on the current state~\cite{tks82}. Hence, the analysis can be quantifying the probability or rate of all safety-related faults: How likely the system is available to meet a demand for service? What is the probability that the system repairs after a failure (e.g., the system conforms to the existent and prominent standards such as the \textit{Safety Integrity Levels} (SILs))?

In order to conduct such analysis, a general approach is modeling and analyzing a probabilistic model of the system (i.e, Markov chains, stochastic processes), in which the algorithm for computing the measures in properties depends on the class of systems being considered and the logic used for specifying the property. Many algorithms with the corresponding mature tools are based on model checking techniques that compute the probability by a numerical approach~\cite{brv04,cg04,rkn04,hwz08}. Timed automata with mature verification tools such as UPPAAL~\cite{lpy05} are used to verify real-time systems. For a variety of probabilistic systems, the most popular modeling formalism is Markov chain or Markov decision processes, for which \textit{Probabilistic Model Checking} (PMC) tools such as PRISM~\cite{hkn06} and MRMC~\cite{khh09} can be used. It is widely used and has been successfully applied to the verification of a range of timed and probabilistic systems. One of the main challenges is the complexity of the algorithms in terms of execution time and memory space due to the size of the state space that tends to grow exponentially, also known as the state space explosion. As a result, the analysis is infeasible. In addition, these tools cannot work directly with the SystemC source code, meaning that a formal model of SystemC model needs to be provided.

An alternative way to evaluate these systems is \textit{Statistical Model Checking} (SMC), a simulation-based approach. Simulation-based approaches produce an approximation of the value to evaluate, based on a finite set of system's executions. Clearly, comparing to the numerical approach, a simulation-based solution does not provide an exact answer. However, users can tune the statistical parameters such as the confidence interval and the confidence, according to the requirements. Simulation-based approaches do not construct all the reachable states of the model-under-verification (MUV), thus they require far less execution time and memory space than numerical approaches. For some real-life systems, they are the only one option~\cite{ykn06} and have shown the advantages over other methods such as PMC~\cite{hwz08,jcj09}.  

Our overall framework weaves the idea of statistical model checking to yield qualitative and quantitative analysis for the probability of a temporal property for SystemC models. The paper makes the following contributions: (i) we propose a framework to verify bounded temporal properties for SystemC models with both timed and probabilistic characteristics. The framework contains two main components: a \textit{monitor} that observes a set of execution traces of the MUV and a statistical model checker implementing a set of hypothesis testing algorithms. We use the similar techniques proposed by Tabakov et al.~\cite{tva10} to automatically generate the monitor. The statistical model checker is implemented as a plugin of the checker Plasma Lab~\cite{bcl13}, in which the properties to be verified are expressed in \textit{Bounded Linear Temporal Logic} (BLTL); (ii) we present a method that allows users to expose a rich set of user-code primitives in form of atomic propositions in BLTL. These propositions help users exposing the state of the SystemC simulation kernel and the full state of the SystemC source code model. In addition, users can define their own fine-grained time resolution that is used to reason on the semantics of the logic expressing the properties rather the boundary of clock cycles in the SystemC simulation; and (iii) we demonstrate our approach through a running example, in which we showcase how our SMC-based verification framework works. We also illustrate the performance of the framework through some experiments.

%% file: systemc.tex
\section{Preliminaries}

\subsection{The SystemC Language}
SystemC is a system-level design framework that can model both hardware and software components. Complex electronic systems and control units can be combined into a single model, to simulate and observe the behavior. In 2005 SystemC became standard as IEEE 1666-2005.

The design process can be parallel with SystemC since it allows blocks implemented at different abstraction levels to run together in the same model. Communication between modules is specified using well-defined interfaces, which allows two modules that conform to the same interface to be swapped seamlessly. Therefore, designers can try alternative approaches early in the design process, before committing to a particular architecture.

SystemC is a library of C$^{++}$ classes that means every SystemC model can be compiled with standard C$^{++}$ compiler and linked with SystemC library to produce an executable specification. SystemC also provides an event-driven mechanisms for simulating parallel execution of the model's processes. The kernel borrows the \textit{delta-cycle} concept from hardware design languages.

\subsubsection{Time Model}
In SystemC, integer values are used as discrete time model. The smallest quantum of time that can be represented is called \textit{time resolution} meaning that any time value smaller than the time resolution will be rounded off. The available time resolutions are femtosecond, picosecond, nanosecond, microsecond, millisecond, and second. SystemC provides functions to set time resolution and declare a time object, for example, the following statements set the time resolution to 10 picosecond and create a time object \begin{tt}t1\end{tt} representing 20 picoseconds:
\begin{lstlisting}[mathescape]
sc_set_time_resolution(10,SC_PS);
sc_time t1(20,SC_PS);
\end{lstlisting}
\subsubsection{Modules}
A SystemC model is composed of \textit{modules}, which define the behavior of the modeled systems. Module data is inaccessible by the other modules of the system unless it is exposed explicitly. Thus, modules can be developed independently and can be reused. The skeleton of a module is given in \lstref{lst:module}:
\begin{lstlisting}[caption=Skeleton code of SystemC module,label=lst:module,mathescape]
SC_MODULE(Name) {
   // prots, processes, internal data, etc

   SC_CTOR(Name) {
      // Body of constructor,
      // Process declaration,
      // Sensitivities, etc.
   }
};
\end{lstlisting}
In general, a module contains:
\begin{itemize}
\item ports which are used to communicate with the environment;
\item processes that represent the functionality of the module;
\item local data and channels to represent the states of module and communication between processes; and
\item hierarchically, other modules.
\end{itemize}
The \begin{tt}SC\_MODULE\end{tt} marco defines a class named \begin{tt}Name\end{tt} and the \begin{tt}SC\_CTOR\end{tt} defines its constructor, which maps designeated methods to \textit{processes} and declares \textit{event sensitives}. A module can be \textit{instantiated} which is similar to the instantiation of class. However, to instantiate a module the user is required to suply a name to the instance. For example, to declare an instance of module \begin{tt}Name\end{tt} named ``xy'', we state:
\begin{lstlisting}[mathescape]
Name xy(``xy'');
\end{lstlisting}
\subsubsection{Interfaces, Ports, and Channels}
In hardware modeling languages, the hardware signal is used as the medium for communication and synchronization between processes. The communication and synchronization are abstracted in SystemC as \textit{interfaces}, \textit{ports}, and \textit{channels} to provide the flexibility. Channels hold and transmit data, and an interface is a ``window'' into a channel that describes the set of operations of the channel. Ports are proxy objects that facilitate access to channels through interfaces.

An interface which is derived from the abstract base class \begin{tt}sc\_interface\end{tt} consists of a set of operations by specifying their \textit{signatures}. We consider a simple interface used with the hardware signal: \begin{tt}sc\_signal\_in\_if<T>\end{tt} which is derived directly from \begin{tt}sc\_interface\end{tt} and is parameterized by data-type \begin{tt}T\end{tt}. It provides a virtual method \begin{tt}read()\end{tt} that returns a constant reference to \begin{tt}T\end{tt}.

A module uses its ports to connect to and communicate with its environment via a channel's interface. Ports can be considered as the pins of a hardware component. A channel has to implement a port's interface to connect to the port. Any specialized port is derived from the port base class \begin{tt}sc\_port\end{tt}.
\begin{lstlisting}[mathescape]
// N is number of channels that can be connected to the port
sc_port<if<ty>,N> p;
\end{lstlisting}
Here we declare a port \begin{tt}p\end{tt}, which can access the number of $N$ channels through the interface \begin{tt}if\end{tt} with type \begin{tt}ty\end{tt}. SystemC provides the following predefined ports, called \textit{signal ports}: \begin{tt}sc\_in\end{tt}, \begin{tt}sc\_out\end{tt}, and \begin{tt}sc\_inout\end{tt} for input, output, and input-output ports. For example:
\begin{lstlisting}[mathescape]
sc_in<int> a;
sc_out<int> b;
\end{lstlisting} 
Here we define an input and an ouput ports named \begin{tt}a\end{tt} and \begin{tt}b\end{tt}, respectively, all of data type \begin{tt}int\end{tt}.

A channel is an implementation  of an interface by providing concrete definitions of all of the interface's operations. Thus, different channels may implement the same interface in different ways. On other hand, a channel can implement more than one interface. Channels provode means fo communication between modules and between processes within a module. The following are several classes of channels in SystemC:
\begin{itemize}
\item A primitive channel does not contain any hierarchy or process and is derived from the base class \begin{tt}sc\_prim\_channel\end{tt}. SystemC contains several built-in channels: \begin{tt}sc\_signal\end{tt}, \begin{tt}sc\_mutex\end{tt}, \begin{tt}sc\_semaphore\end{tt}, and \begin{tt}sc\_fifo\end{tt}.
\item A hierarchical channel can have a structure, contain processes and access directly other channels. All hiercarchical channels are derived from the base class \begin{tt}sc\_channel\end{tt} that is just a redefinition of the class \begin{tt}sc\_module\end{tt}. Thus from a language point of view a hierarchical channel is nothing but a module. 
\end{itemize}
\subsubsection{Processes}
 Processes which provide the mechanism for simulating concurrent behavior are basic units of functionality. A process must be contained in a module and declared to be a process in the module's constructor. There are two kinds of processes: \textit{method process} (with macro \begin{tt}SC\_METHOD\end{tt}) and \textit{thread process} (with macro \begin{tt}SC\_THREAD\end{tt}).

When triggered, a method process always executes its body until the return. That means it only returns the control to the kernel when it is at the end of its body. A thread process, on the other hand, may have its execution suspended by calling the library function \begin{tt}wait()\end{tt} or any of its variants. All local variables and the point of suspension are saved. When the execution is resummed, it will continue from that point, rather than from the beginning of the process. Thus, unlike method processes, a thread process implicitly keeps its state of execution. This feature makes thread process more expressive than method process, for example, by means of \begin{tt}wait\end{tt} statements multicycle behavior may be easily described by thread process, but would require more effort with method process.
\subsubsection{Events}
An event is an object C$^{++}$ of class \begin{tt}sc\_event\end{tt}, that determines whether and when a process's execution should be triggered or resumed. By default, SystemC defines for each \begin{tt}sc\_signal\end{tt} an associated event \begin{tt}value\_changed\_event()\end{tt}, that is notified whenever the value of the signal is written or modified. The effect of the notification (by calling \begin{tt}e.notify()\end{tt}) of \begin{tt}e\end{tt} causes all processes that are sensitive to it (or use \begin{tt}wait(e)\end{tt}) to be triggered or resumed. The notification can have different effects, depending on its argument:
\begin{itemize}
\item \begin{tt}notify()\end{tt} without arguments makes \textit{immediate notification} and puts all processes that are sensitive to the event to the pool of runnable processes before the return of the function call \begin{tt}notify()\end{tt}.
\item \begin{tt}notify()\end{tt} with arguments as zero time unit (e.g., \begin{tt}SC\_ZERO\_TIME\end{tt}) delays the effect of the event notification until all currently triggered processes have finished executing. The simulation clock does not advance during this delay. It is called \textit{delta-delayed} notification.
\item \begin{tt}notify()\end{tt} with arguments as non-zero time units delays the effect of the notification by the number of time units. The argument value is added to the simulation clock, and the event is put in a queue. It is called \textit{time-delayed} notification.
\end{itemize}
All event notifications are pending, they can be \textit{canceled}, which removes any pending effect of the event. A process can wait for an event in some bounded of time. For example, \begin{tt}wait(2,SC\_SEC,e)\end{tt} resumes the execution after 2 seconds of simulation time if \begin{tt}e\end{tt} is notified earlier.  
\subsubsection{Sensitivity}
A module can be sensitive to events which is declared via the \begin{tt}<<\end{tt} operator as follows:
\begin{lstlisting}[mathescape]
// special attribute of module named sensitive
sensitive << ``event1'' << ``event2'';
\end{lstlisting}
If the list of events remains the same throughout simulation, it is called a \textit{static sensitivity list}. Otherwise, it is a \textit{dynamic sensitivity list}. That is, during  simulation a thread process may suspend itself and designate a specific event \begin{tt}e\end{tt} as its current waiting event. Then, only this event can resume the execution of the process (the static sensitivity list is ignored). A process can wait for an event, composite events, or for a time:
\begin{lstlisting}[mathescape]
wait(e);
wait(e1 & e2 & e3);
wait(e1 | e2 | e3);
wait(10,SC_NS);
\end{lstlisting}
\subsubsection{Simulation Kernel}
\label{subsec:simulationsemantics}
The SystemC simulation kernel is an event-driven simulation. The structural information is represented by the modules and ports. Only one thread is dispatched by the scheduler to run at a time point, and the scheduler is non-preemptive, that is, the running process returns control to the kernel only when it finishes executing or explicitly suspends itself by calling \begin{tt}wait()\end{tt}.

Like hardware modeling languages, the SystemC scheduler supports the notion of delta-cycles \cite{lsu93}. A delta-cycle lasts for an infinitesimal amount of time and is used to impose a partial order of simultaneous actions which interprets zero-delay semantics. Thus, the simulation time is not advanced when the scheduler processes a delta-cycle. During a delta-cycle, the scheduler executes actions in two phases: the \textit{evaluate} and the \textit{update} phases. The simulation semantics of the SystemC scheduler is presented as follows:
\begin{enumerate}
\item \textit{Initialize}. During the initialization, each process is executed once unless it is turned off by calling \begin{tt}dont\_initialize()\end{tt}, or until a synchronization point (i.e., a \begin{tt}wait\end{tt}) is reached. The order in which these processes are executed is unspecified.
\item \textit{Evaluate}. The kernel starts a delta-cycle and run all processes that are ready to run one at a time. In this same phase a process can be made ready to run by an event notification.
\item \textit{Update}. Execute any pending calls to \begin{tt}update()\end{tt} resulting from calls to \begin{tt}request\_update()\end{tt} in the evaluate phase. Note that a primitive channel uses \begin{tt}request\_update()\end{tt} to have the kernel call its \begin{tt}update()\end{tt} function after the execution of processes.
\item The kernel enters the delta notifcation phase where notified events trigger their dependent processes. Note that immediate notifications may make new processes runable during step 2. If so the kernel loops back to step 2 and starts another evaluation phase and a new delta-cycle. It does not advance simulation time.
\item If there are no more runable processes, the kernel advances simulation time to the earliest pending timed notification. All processes sensitive to this event are triggered and the kernel loops back to step 2 and starts a new delta-cycle. This process is finised when all processes have terminated or the specified simulation time is passed. 
\end{enumerate}
The simulation semantics can be represented by the pseudo code in \lstref{lst:simulationsemantics}.
\begin{lstlisting}[caption=Simulation Semantics of SystemC,label=lst:simulationsemantics,mathescape]
PC // All primitive channels
P // All processes
R $\leftarrow$ $\emptyset$ // Set of runnable processes
D $\leftarrow$ $\emptyset$ // Set of pending delta notifications
U $\leftarrow$ $\emptyset$ // Set of update requests
T $\leftarrow$ $\emptyset$ // Set of pending timed notifications
// Start elaboration: collect all update requests in U
for all chan $\in$ U do
   run chan.update()
end for
for all p $\in$ P do
   if p is initialized and p is not clocked thread then
      R $\leftarrow$ R $\cup$ p // Make p runnable
   end if
end for
for all p $\in$ P do
   if p is triggered by an event in D then 
      R $\leftarrow$ R $\cup$ p
   end if
end for // End of initialization phase

repeat
   while R $\neq$ $\emptyset$ do // New delta-cycle begins
      for all r $\in$ R do // Evaluation phase
         R $\leftarrow$ R $\setminus$ r
         run r until it calls wait() or returns
      end for
      for all chan $\in$ U do // Update phase
         run chan.update()
      end for
      for all p $\in$ P dp // Delta notification phase
         if p is triggered by an event in D then
            R $\leftarrow$ R $\cup$ p // Make p runnable
         end if
      end for // End of delta-cycle
   end while
   
   if T $\neq$ $\emptyset$ then
      Advance the simulation clock to the earliest timed delay t
      T $\leftarrow$ T $\setminus$ t      
      for all p $\in$ P do // Timed notification phase
         if t triggers p then
            R $\leftarrow$ R $\cup$ p // Make p runnable
         end if
      end for
   end if
until end of simulation
\end{lstlisting}
\subsection{Example: Producer-consumer Model}
\label{subsec:producer-consumer}
\begin{figure}[ht]
\begin{center}
\includegraphics[width=0.90\textwidth]{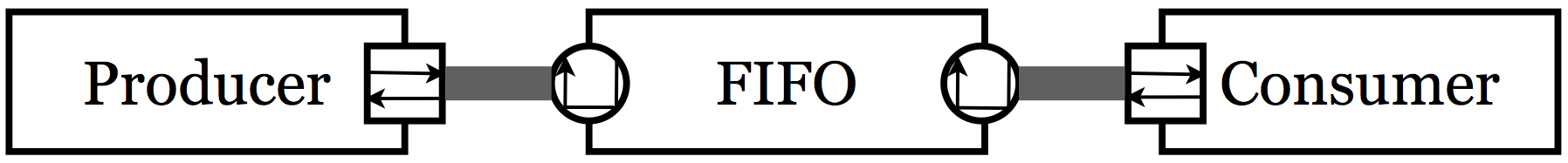}
\caption{Producer and consumer example}
\label{fig:fifo}
\end{center}
\end{figure}

We will use a simple case study with a FIFO channel as a running example (see \figref{fig:fifo} with the graphical notations in \cite{glm02}). 
This example shows the communication between modules throught a shared channel. The model consists of two modules \begin{tt}Producer\end{tt} and \begin{tt}Consumer\end{tt} that communicate via a fixed length FIFO. It demonstrates how construct these modules and the communication channel using \begin{tt}sc\_interface\end{tt} and \begin{tt}sc\_channel\end{tt} based classes. In this example, we will build a simple channel for character writing and reading. \lstref{lst:fifointerfaces} shows a character interface definition. We can see that our interfaces have to derived from the based class \begin{tt}sc\_interface\end{tt} and have only pure virtual methods.
\subsubsection{Interfaces}
\begin{lstlisting}[caption=The fifo\_if.h,label=lst:fifointerfaces,mathescape]
#ifndef FIFO_IF
#define FIFO_IF
#include <systemc.h>

class fifo_write_if : virtual public sc_interface {
public:
   virtual void fifo_write(char) = 0;
   virtual void fifo_reset() = 0;   
};

class fifo_read_if : virtual public sc_interface {
public:
   virtual void fifo_read(char&) = 0;
   virtual int fifo_num_available() = 0;
};

#endif
\end{lstlisting}
\subsubsection{Modules}
The \begin{tt}Producer\end{tt} writes a character to the FIFO with the probability of $p_1 \in [0,1]$ and the \begin{tt}Consumer\end{tt} reads a character from the FIFO with the probability $p_2 \in [0, 1]$ for every nanosecond. We model the probabilities of writing and reading with the Bernoulli distributions with probabilities $p_1$ and $p_2$ respectively from GNU Scientific Library (GSL)~\cite{gsl}. The following listings implement the specification of the \begin{tt}Producer\end{tt} and the \begin{tt}Consumer\end{tt}.
\begin{lstlisting}[caption=The consumer.h,label=lst:consumerh,mathescape]
#ifndef CONSUMER_H
#define CONSUMER_H

#include <systemc.h>
#include <tlm.h>
#include "fifo.cpp"
#include "utils.h"
#include <gsl/gsl_rng.h>
#include <gsl/gsl_randist.h>
#include <gsl/gsl_cdf.h>

SC_MODULE(Consumer) {
   SC_HAS_PROCESS(Consumer);
public:
   // Definitions of ports
   sc_port<fifo_read_if> in; // input port
   // Constructor
   Consumer(sc_module_name name, int c_init, gsl_rng *rnd);
   // Destructor
   ~Consumer() {};
   // Definition of processes
   void main();
   // Reading function
   void receive(char &c);

private:
   // Reading character in ASCII
   int c_int;
   gsl_rng *r;
};

#endif
\end{lstlisting}
\begin{lstlisting}[caption=The consumer.cpp,label=lst:consumercpp,mathescape]
#include "consumer.h"

Consumer::Consumer(sc_module_name name, int c_init, gsl_rng *rnd) {
   c_int = c_init;
   r = rnd; // random generator
 
   SC_THREAD(main);
}

void Consumer::receive(char &c) {
   in->fifo_read(c);
   c_int = c;
}

void Consumer::main() {
   char c;
   while (true) {
      // use the Bernoulli distribution in GSL
      int b = get_bernoulli(r,0.90);
      if (b) {
         receive(c);
         std::cout << c << "(" << sc_time_stamp() << ")";
      }
   
      wait(1,SC_NS); // waits for 1 nanosecond
   }  
}
\end{lstlisting}
\begin{lstlisting}[caption=The producer.h,label=lst:producerh,mathescape]
#ifndef PRODUCER_H
#define PRODUCER_H

#include <systemc.h>
#include <tlm.h>
#include "fifo.cpp"
#include "utils.h"
#include <gsl/gsl_rng.h>
#include <gsl/gsl_randist.h>
#include <gsl/gsl_cdf.h>

SC_MODULE(Producer) {
   SC_HAS_PROCESS(Producer);
public:
   // Definitions of ports
   sc_port<fifo_write_if> out; // output port
   // Constructor
   Producer(sc_module_name name, int c_init, gsl_rng *rnd);
   // Destructor
   ~Producer() {};
   // Definition of processes
   void main();
   // Writing function
   void send(char c);

private:
   int c_int;
   gsl_rng *r;
};

#endif
\end{lstlisting}
\begin{lstlisting}[caption=The producer.cpp,label=lst:producercpp,mathescape]
#include "producer.h"

Producer::Producer(sc_module_name name, int c_init, gsl_rng *rnd) {
   c_int = c_init;
   r = rnd; // random generator

   SC_THREAD(main);
}

void Producer::send(char c) {
   out->fifo_write(c);
   c_int = c;
}

void Producer::main() {
   const char* str = "&abcdefgh@";
   const char* p = str;
   while (true) {
      int b = get_bernoulli(r,0.90);
      if (b) {
         send(*p);
         p++;
         if (!*p) {
            p = str;
         }
      }
      wait(1,SC_NS); // waits for 1 nanosecond
   }  
}
\end{lstlisting}
The producer has one thread which runs infinitely to write a character into the FIFO that it connects to via the output port using the interface \begin{tt}fifo\_write\_if\end{tt} (line 16 in \begin{tt}producer.h\end{tt}). The producer's process suspends itself explicitly by calling \begin{tt}wait()\end{tt} (line 27 in \begin{tt}producer.cpp\end{tt}). The implementation of the consumer is in the similar way.
\subsubsection{Channel}
Now we implement the FIFO that is derived from \begin{tt}fifo\_write\_if\end{tt} and \begin{tt}fifo\_read\_if\end{tt} as in \lstref{lst:fifo}.
\begin{lstlisting}[caption=The fifo.cpp,label=lst:fifo,mathescape]
#ifndef BASE_CHANNEL_H
#define BASE_CHANNEL_H
#include <systemc.h>
#include "fifo_if.h"

class Fifo : public sc_channel, public fifo_write_if, public fifo_read_if {
private:
   enum e {max = 10}; // capacity of the fifo
   char data[max];
   int num_elements, first;
   sc_event write_event, read_event;

public:
   Fifo(sc_module_name name) : sc_channel(name), num_elements(0), first(0) {}
   
   void fifo_write(char c) {
      if (num_elements == max) {
         wait(read_event);
      }

      data[(first + num_elements) % max] = c;
      ++num_elements;
      write_event.notify();
   }

   void fifo_read(char &c) {
      if (num_elements ==  0) {
         wait(write_event);
      }
      
      c = data[first];
      --num_elements;
      first = (first + 1) % max;
      read_event.notify();
   }

   void fifo_reset() {
      num_elements = 0;
      first = 0;
   }

   int fifo_num_available() {
      return num_elements;
   }
};

#endif
\end{lstlisting}
Since the FIFO is bounded capacity meaning that it may be full when the producer tries to write a character, or it may be empty when the consumer attempts to read a character. Thus, the implementation of the FIFO must handle this situation using the \textit{blocking} \begin{tt}read()\end{tt} and \begin{tt}write()\end{tt} operations. The channel \begin{tt}fifo\end{tt} has the local variabels to store the available characters, the positions of the next character to read and to write.

The \begin{tt}read()\end{tt} operation first checks the number of available characters. If the fifo is empty, the operation suspends execution with a call to \begin{tt}wait(write\_event)\end{tt} (line 28) and the execution is resumed only when the \begin{tt}write\_event\end{tt} is notified. As soon as a character is read from the fifo, the event \begin{tt}read\_event\end{tt} will be notified (line 34).

The \begin{tt}write()\end{tt} operation is implemented similarly. It checks that whether the fifo is full or not. If the fifo is full, the operation suspends execution with a call to \begin{tt}wait(read\_event)\end{tt} (line 18) and the execution is resumed only when the \begin{tt}read\_event\end{tt} is notified. As soon as a character is writtend to the fifo, the event \begin{tt}write\_event\end{tt} will be notified (line 23).

The FIFO channel is designed to ensure that all data is reliably delivered despite the varying rates of production and consumption. The channel uses an event notification hanshake protocol for both the input and output. It uses a circular buffer implemented within a static array to store and retrieve the items within the FIFO. We assume that the sizes of the messages and the FIFO buffer are fixed. Hence, it is obvious that the time required to transfer completely a message, or message \textit{latency}, depends on the production and consumption rates, the FIFO buffer size, the message size, and the probabilities of successful writing and reading. 
\subsubsection{Binding and Simulation}
We now bind the modules of the producer and consumer with the fifo channel via the output and input ports. One can write the binding code in a separated module or inside the \begin{tt}sc\_main()\end{tt} function that is the entry of the executable SystemC model. The following listing presents an example of binding and simulation of the producer-consumer model.
\begin{lstlisting}[caption=The main.cpp,label=lst:main,mathescape]
#include <time.h>
#include "fifo.cpp"
#include "consumer.h"
#include "producer.h"

#include <gsl/gsl_rng.h>
#include <gsl/gsl_randist.h>
#include <gsl/gsl_cdf.h>
// The monitor generated by MAG
#include "monitor.h"

int sc_main(int argc, char *argv[]) {
   // random generator in GSL
   const gsl_rng_type *T;
   gsl_rng *r;
   gsl_rng_env_setup();
   T = gsl_rng_default;
   r = gsl_rng_alloc(T);
   // seed the generator
   srand(time(NULL));
   gsl_rng_set(r,random());
 
   sc_set_time_resolution(1,SC_NS); // time unit
   Fifo afifo("fifo"); // create a channel fifo
   Producer prod("producer",-1,r);
   Consumer cons("consumer",-1,r);
   prod.out(afifo); // the producer binding
   cons.in(afifo);  // the consumer binding
   // the observer for Instrumented model
   mon_observer* obs = local_observer::createInstance(1, 
                       &cons, 
                       &prod);
   sc_start();
   gsl_rng_free (r); // release the generator
   return 0;
}
\end{lstlisting}

If we compile and run the file \begin{tt}main.cpp\end{tt}, some of the possible outputs of the model are given as follows:
\begin{lstlisting}[caption=Simulation Outputs,label=lst:main]
(&, 0) (a, 1)        (b, 3)                     (c, 8) (d, 9)
                            (&, 4)
(&, 0)       (a, 2)  (b, 3) (c, 4)                     (d, 9) 
\end{lstlisting}
$(x, i)$ means that the consumer reads a character ``x'' at the $(i+1)th$ nanosecond.

\subsection{Statistical Model Checking}
We first recall the syntax and semantics of BLTL~\cite{sva05}, an extension of Linear Temporal Logic (LTL) with time bounds on temporal operators. A formula $\varphi$ is defined over a set of atomic propositions $AP$ as in LTL. A BLTL formula is defined by the grammar $\varphi ::= true | false | p \in AP | \varphi_1 \wedge \varphi_2 | \neg \varphi | \varphi_1 \; U_{\leq T} \; \varphi_2$, where the time bound $T$ is an amount of time or a number of states in the execution trace. The temporal modalities $F$ (the ``eventually'', sometimes in the future) and $G$ (the ``always'', from now on forever) can be derived from the ``until'' $U$ as follows.
\begin{displaymath}
F_{\leq T} \; \varphi = true \; U_{\leq T} \; \varphi \text{ and } G_{\leq T} \; \varphi = \neg F_{\leq T} \; \neg \varphi
\end{displaymath}
The semantics of BLTL is defined w.r.t execution traces of the model $\mathcal{M}$. Let 
\begin{displaymath}
\omega = (s_0,t_0)(s_1,t_1)...(s_{N-1},t_{N-1}), N \in \mathbb{N}
\end{displaymath} 
be an execution trace of $\mathcal{M}$, $\omega_k$ and $\omega^k$ be the prefix and suffix of $\omega$ respectively. We denote the fact that $\omega$ satisfies the BLTL formula $\varphi$ by $\omega \models \varphi$. 
\begin{itemize}
\item $\omega^k \models true$ and $\omega^k \not \models false$
\item $\omega^{k} \models p, p \in AP$ iff $p \in L(s_k)$, where $L(s_k)$ is the set of atomic propositions which are $true$ in state $s_k$ 
\item $\omega^{k} \models \varphi_1 \wedge \varphi_2$ iff $\omega^k \models \varphi_1$ and $\omega^k \models \varphi_2$
\item $\omega^k \models \neg \varphi$ iff $\omega^k \not \models \varphi$
\item $\omega^k \models \varphi_1 \; U_{\leq T} \; \varphi_2$ iff there exists $i \in \mathbb{N}$ such that $\omega^{k+i} \models \varphi_2$, $\Sigma_{0 < j \leq i}(t_{k+j} - t_{k+j-1}) \leq T$, and for each $0 \leq j < i, \omega^{k+j} \models \varphi_1$
\end{itemize}

Let $\mathcal{M}$ be the formal model of the MUV (i.e., a stochastic process) and $\varphi$ be a property expressed as a BLTL formula. BLTL ensures that the satisfaction of a formula by a trace can be decided in a finite number of steps. The statistical model checking~\cite{ldb10} problem consists in answering the following questions: (i) Is the probability that $\mathcal{M}$ satisfies $\varphi$ greater or equal to a threshold $\theta$ with a specific level of statistical confidence (\textit{qualitative analysis})? (ii) What is the probability that $\mathcal{M}$ satisfies $\varphi$ with a specific level of statistical confidence (\textit{quantitative analysis})? They are denoted by $\mathcal{M} \models Pr(\varphi)$ and $\mathcal{M} \models Pr_{\geq \theta}(\varphi)$, respectively. Many statistical model checker are implemented~\cite{yhl05,bcl13} that have shown their advantages over other methods such as PMC on several case studies.

This is done by associating each execution trace of $\mathcal{M}$ with a discrete random Bernoulli variable $B_i$, in which the outcome for $B_i$, denoted by $b_i$, is $1$ if the trace satisfies $\varphi$ and $0$ otherwise. The predominant statistical method for verifying $\mathcal{M} \models Pr_{\geq \theta}(\varphi)$ is based on \textit{hypothesis testing}. Let $p = Pr(\varphi)$, to determine whether $p \geq \theta$, we test the hypothesis $H_0: p \geq p_0 = \theta + \delta$ against the alternative hypothesis $H_1: p \leq p_1 = \theta - \delta$ based on the observations of $B_i$. The size of \textit{indifference region} is defined by $p_0 - p_1$. If we take acceptance of $H_0$ to mean acceptance of $Pr_{\geq \theta}(\varphi)$ as true and acceptance of $H_1$ to mean rejection of $Pr_{\geq \theta}(\varphi)$ as false, then we can use \textit{acceptance sampling} (e.g., Younes in~\cite{you05} has proposed two solutions, called \textit{single sampling plan} and \textit{sequential probability ratio test}) to verify $Pr_{\geq \theta}(\varphi)$. An acceptance sampling test with \textit{strength} $(\alpha,\beta)$ guarantees that $H_1$ is accepted with probability at most $\alpha$ when $H_0$ holds and $H_0$ is accepted with probability at most $\beta$ when $H_1$ holds, called a Type-I error and Type-II error, respectively. 

To answer the quantitative question, $\mathcal{M} \models Pr(\varphi)$, an alternative statistical method, based on \textit{estimation} instead of hypothesis testing, has been developed. For instance, the probability estimations are based on results derived by Chernoff and Hoeffding bounds~\cite{hoe63}. This approach uses $n$ observations $b_1,...,b_n$ to compute an approximation of $p$: $\tilde{p} = \frac{1}{n}\Sigma^{n}_{i=1}b_i$. The approximation satisifies that $Pr[|\tilde{p} - p| < \delta] \geq 1 - \alpha$. Based on the theorem of Hoeffding, the number of observations which is determined from the absolute error $\delta$ and the confidence $1 - \alpha$ is $n = \lceil \frac{1}{2\delta^2}log\frac{2}{\alpha} \rceil$.

Although SMC can only provide approximate results with a user-specified level of statistical confidence, it is compensated for by its better scalability and resource consumption. Since the models to be analyzed are often approximately known, an approximate result in the analysis of desired properties within specific bounds is quite acceptable. SMC has recently been applied in a wide range of research areas including software engineering (e.g., verification of critical embedded systems)~\cite{hwz08}, system biology, or medical area~\cite{jcj09}.

%% file: trace.tex
\section{SMC for SystemC Models}
\label{sec:smcsystemc}
In order to apply SMC for SystemC models which exhibit probabilistic characteristics, this section presents the definitions of state and execution trace of SystemC models. This section also shows that the operational semantics of this class of SystemC models is considered as stochastic processes.
\subsection{SystemC Model State}
Temporal logic formulas are interpreted over execution traces and traditionally a trace has been defined as a sequence of states in the execution of a model. Therefore before we can define an execution trace we need a precise definition of the state of a SystemC model simulation. We are inspired by the definition of system state in~\cite{tva10}, which consists of the state of the simulation kernel and the state of the SystemC model. We consider the external libraries as black boxes, meaning that their states are not exposed. 

The state of the kernel contains the information about the current phase of the simulation (i.e., delta-cycle notification, simulation-cycle simulation) and the SystemC events notified during the execution of the model. The state of the SystemC model is the full state of the C++ code of all modules in the model, which includes the values of the module attributes, the location of the program counter (i.e., a particular statement is reached during the execution of the model, the function calls), the call stack including the function execution, function parameters and return values, and the status of the module processes (i.e., suppended, runnable). We use $V = \{v_0,...,v_{n-1}\}$ to denote the finite set of variables of primitive type (e.g, usual scalar or enumerated type in C/C++) whose value domain $\mathbb{D}_X$ represents the states of a SystemC model. 

We consider here some examples about states of the simulation kernel and the SystemC model. Assume that a SystemC model has an event named $e$, then the model state can contain information such as the kernel is at the end of simulation-cycle notification phase and the event $e$ is notified. Consider the running example again, a state can consist of the information about the characters received by the consumer, represented by the variable $c\_read$. It also contains the information about the location of the program counter right before and after a call of the function \textit{send()} in the module \textit{Producer} that are represented by two Boolean variables \textit{send\_start} and \textit{send\_done}, respectively, meaning that they hold the value \textit{true} immediately before and after a call of the function \textit{send()}. Another example, we consider a module that consists several statements at different locations in the source code, in which these statements contain the devision operator ``/'' followed by zero or more spaces and the variable ``\textit{a}'' (e.g., the statement \textit{y = (x + 1) / a}). Then, a Boolean variable which holds the value \textit{true} right before the execution of all such statements can be used as a part of the states.

We have discussed so far the state of a SystemC model execution. It remains to discuss how the semantics of the temporal operators is interpreted over the states in the execution of the model. That means how the states are sampled in order to make the transition from one state to another state. The following definition gives the concept of \textit{temporal resolution}, in which the states are evaluated only at instances in which the temporal resolution holds. It allows the user to set granularity of time. 
\begin{definition}[Temporal resolution]
A temporal resolution $\mathcal{T}_r$ is a finite set of Boolean expressions defined over $V$ which specifies when the set of variables $V$ is evaluated.
\end{definition}
Temporal resolution can be used to define a more fine-grained model of time than a coarse-grained one provided by a cycle-based simulation. We call the expressions in $\mathcal{T}_r$ \textit{temporal events}. Whenever a temporal event is satisfied or the temporal event occurs, $V$ is sampled. For example, in the producer and consumer model, assume that we want the satisfaction of the underlying BLTL $\varphi$ to be checked whenever at the end of simulation-cycle notification or immediately after the event \textit{write\_event} is notified during a run of the model. Hence, we can define a temporal resolution as the following set $\mathcal{T}_r =$ $\{end\_sc, we\_notified\}$, where \textit{end\_sc} and \textit{we\_notified} are Boolean expressions that have the value \textit{true} whenever the kernel phase is at the end of the simulation-cycle notification and the event \textit{write\_event} notified, respectively.

We denote the set of occurrences of temporal events from $\mathcal{T}_r$ along an execution of a SystemC model by $\mathcal{T}^{s}_r$, called a \textit{temporal resolution set}. The value of a variable $v \in V$ at an event occurrence $e_c \in \mathcal{T}^{s}_{r}$ is defined by a mapping $\xi^{v}_{val}: \mathcal{T}^{s}_{r} \rightarrow \mathbb{D}_{v}$, where $\mathbb{D}_{v}$ is the value domain of $v$. Hence, the state of the SystemC model at $e_c$ is defined by a tuple $(\xi^{v_0}_{val},...,\xi^{v_{n-1}}_{val})$.

A mapping $\xi_t: \mathcal{T}^{s}_{r} \rightarrow \mathcal{T}$ is called a \textit{time event} that identifies the simulation time at each occurrence of an event from the temporal resolution. Hence, the set of time points, called \textit{time tag}, which corresponds to a temporal resolution set $\mathcal{T}^{s}_{r} = \{e_{c_0},...,e_{c_{N-1}}\}, N \in \mathbb{N}$, is given as follows.
\begin{definition}[Time tag]
Given a temporal resolution set $\mathcal{T}^{s}_{r}$, the \textit{time tag} $\mathcal{T}$ corresponding to $\mathcal{T}^{s}_{r}$ is a finite or infinite set of non-negative reals $\{t_0,t_1,...,t_{N-1}\}$, where $t_{i+1} - t_i = \delta t_i \in \mathbb{R}_{\geq 0}, t_i = \xi_t(e_{c_i})$.
\end{definition}
\subsection{Model and Execution Trace}
A SystemC model can be viewed as a hierarchical network of parallel communicating processes. Hence, the execution of a SystemC model is an alternation of the control between the model's processes, the external libraries and the kernel process. The execution of the processes is supervised by the kernel process to concurrently update new values for the signals and variables w.r.t the cycle-based simulation. For example, given a set of runnable processes in a simulation-cycle, the kernel chooses one of them to execute first in a non-deterministic manner as described in the prior section.

Let $V$ be the set of variables whose values represent the states of a SystemC model. The values of variables in $V$ are determined by a given probability distribution (i.e., production from all probability distributions used in the model). Given a temporal resolution $\mathcal{T}_r$ and its corresponding temporal resolution set along an execution of the model $\mathcal{T}^{s}_{r} = \{e_{c_0},...,e_{c_{N-1}}\}, N \in \mathbb{N}$, the evaluation of $V$ at the event occurrence $e_{c_i}$ is defined by the tuple $(\xi^{v_0}_{val},...,\xi^{v_{n-1}}_{val})$, or a state of the model at $e_{c_i}$, denoted by $V(e_{c_i}) = (V(e_{c_i})(v_0),V(e_{c_i})(v_1),...,V(e_{c_i})(v_{n-1}))$, where $V(e_{c_i})(v_k) = \xi^{v_k}_{val}(e_{c_i})$ with $k = 0,...,n-1$ is the value of the variable $v_k$ at $e_{c_i}$. We denote the set of all possible evaluations by $V_{\mathcal{T}^{s}_{r}} \subseteq \mathbb{D}_V$, called the \textit{state space} of the random variables in $V$. State changes are observed only at the moments of event occurrences. Hence, the operational semantics of a SystemC model is represented by a \textit{stochastic process} $\{(V(e_{c_i}),\xi_t(e_{c_i})), e_{c_i} \in \mathcal{T}^{s}_{r}\}_{i \in \mathbb{N}}$, taking values in $V_{\mathcal{T}^{s}_{r}} \times \mathbb{R}_{\geq 0}$ and indexed by the parameter $e_{c_i}$, which are event occurrences in the temporal resolution set $\mathcal{T}^{s}_{r}$. An execution trace is a realization of the stochastic process is given as follows.
\begin{definition}[Execution trace]
An execution trace of a SystemC model corresponding to a temporal resolution set $\mathcal{T}^{s}_{r} = \{e_{c_0},...,e_{c_{N-1}}\}, N \in \mathbb{N}$ is a sequence of states and event occurrence times, denoted by $\omega = (s_0,t_0)...(s_{N-1},t_{N-1})$, such that for each $i \in 0,...,N-1$, $s_i = V(e_{c_i})$ and $t_i = \xi_t(e_{c_i})$. 
\end{definition}
$N$ is the length (finite or infinite) of the execution, also denoted by $|\omega|$. Let $V' \subseteq V$, the \textit{projection} of $\omega$ on $V'$, denoted by $\omega \downarrow_{V'}$, is an execution trace such that $|\omega \downarrow_{V'}| = |\omega|$ and $\forall v \in V'$, $\forall e_c \in \mathcal{T}^{s}_{r}$, $V'(e_c)(v) = V(e_c)(v)$.
\subsection{Expressing Properties}
\label{sec:bltl}
Our approach allows users to refer to a rich set of atomic propositions $AP$ which is defined over the set of variables $V$ as previously mentioned. These propositions abstract the states of a SystemC model and evaluate to either \textit{true} or \textit{false} in such a state. The implementation provides a mechanism that allows users to declare $V$ in order to define the set of propositions $AP$ without requiring users to write the monitoring code or to write aspect-oriented programming advices manually.

Users declare these variables via a high-level language in a configuration file as the input of our tool. For instance, refering to the producer and consumer model, the declaration \textit{location send\_start ``\%Producer::send()'':call} declares a Boolean variable \textit{send\_start} that holds the value \textit{true} immediately before the execution of the model reaches a call site of the function \textit{send()} in the module \textit{Producer}. The characters received by the consumer which is represented by the variable $c\_read$ can be declared as \textit{attribute pnt\_con$\rightarrow$c\_int c\_read}, where \textit{pnt\_con} is a pointer to the \textit{Consumer} object and \textit{c\_int} is an attribute of the \textit{Consumer} moudle representing the received character. The detailed syntax can be found in the tool manual\footnote{\url{https://project.inria.fr/plasma-lab/documentation/tutorial/mag_manual/}}.

$AP$ are predicates defined over the set of variables $V$. Using these predicates, users can define temporal properties related to the states of the kernel and the SystemC model. Recall the considered property of the running example, the predicates which are defined over the variable \textit{c\_read} are $c\_read = '\&'$ and $c\_read = '@'$. Another example, assume that we want to answer the following question: \textit{``Over a period of $T$ time units, is the probability that the number of elements in the FIFO buffer in between $n_1$ and $n_2$ is greater or equal to $\theta$ with the confidence $\alpha$?''}. The predicates need to be defined in order to construct the underlying BLTL formula are $n_1 \leq n_{elements}$ and $n_{elements} \leq n_2$, where $n_{elements}$ is an integer variable that represents the current number of elements in the FIFO buffer (it capptures the value of the \textit{num\_elements} attribute in the \textit{Fifo} module). Then, the property can be translated in BLTL with the operator ``always'' as follows. The input which is given to the checker is $Pr_{\geq \theta}(\varphi)$ along with the confidence $\alpha$.
\begin{displaymath}
\varphi = G_{\leq T}((n_1 \leq n_{elements}) \; \& \; (n_{elements} \leq n_2))
\end{displaymath}

%% file: implementation.tex
\section{Implementation}
\label{sec:imp_exp}
\begin{figure}[ht]
\begin{center}
\includegraphics[width=0.90\textwidth]{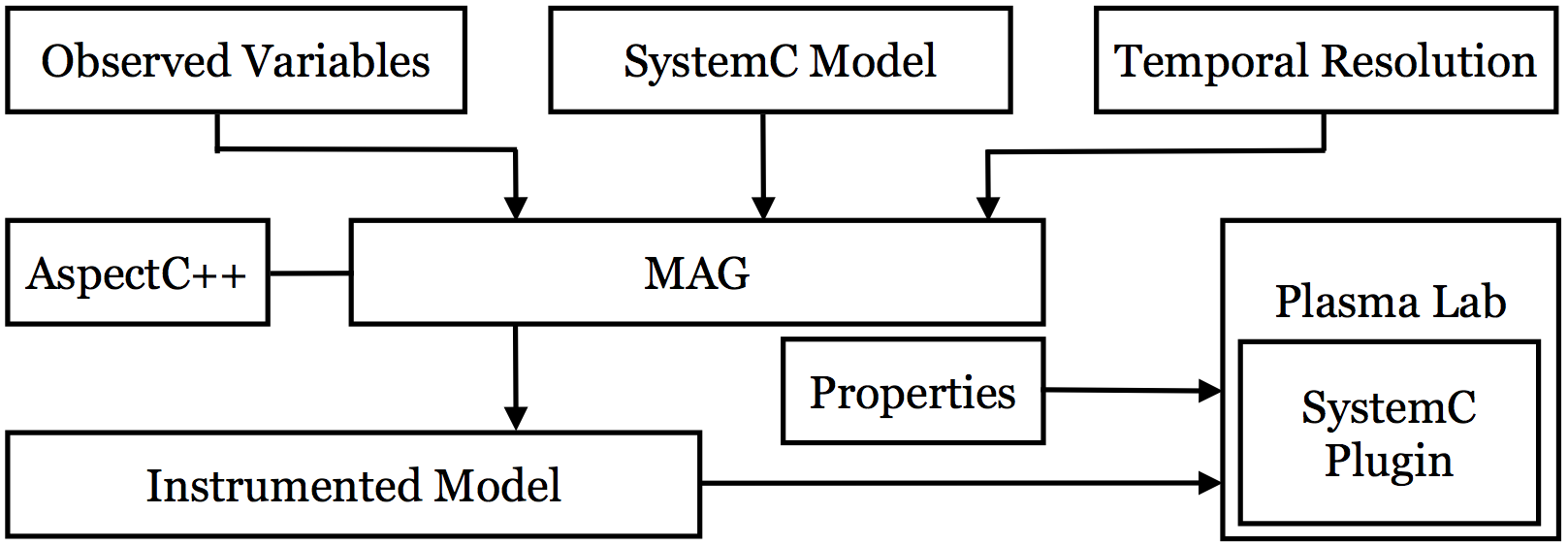}
\caption{The framework's flow}
\label{fig:architecture}
\end{center}
\end{figure}

We have implemented a SMC-based verification tool that contains two main components: a \textit{monitor and aspect-advice generator} (MAG) and a \textit{statistical model checker} (SystemC Plugin). The tool whose flow is depicted in \figref{fig:architecture} can be considered as a static runtime verification tool for probabilistic temporal properties. 
\subsection{MAG and SystemC Plugin}
In principle, the full state can be observed during the simulation of the model. In practice, however, users define a set of variables of interest, according to the properties that the users want to verify, called \textit{observed variables}, and only these variables appear in the states of an execution trace. 
Given a SystemC model, we use $V_{obs} \subseteq V$ to denote the set of variables, called \textit{observed variables}, to expose the states of the SystemC model. Then, the observed execution traces of the model are the projections of the execution traces on $V_{obs}$, meaning that for every execution trace $\omega$, the corresponding observed execution trace is $\omega \downarrow_{V_{obs}}$. In the following, when we mention about execution traces, we mean observed execution traces.

The implementation of MAG allows users to define a set of observed variables that is used with a temporal resolution to generate a monitor. The implementation based on the techniques in~\cite{tva10}, in which a monitor and an aspect-advice file are generated in order to automatically instrument the SystemC model with the help of AspectC++~\cite{gss01} and establish a communication between the generated monitor and the instrumented model. The monitor evaluates the set of observed variables at every time point at which an event of the temporal resolution occurs during the SystemC model simulation to produce a new state. 

For example, the variable $c\_read$ which observes the received character by the consumer (the private attribute $c\_int$ in the module \textit{Consumer}) at the end of simulation-cycle notification, is implemented by generating a monitor and instrumenting the module \textit{Consumer} to establish a communication between them as follows. The module \textit{Consumer} is instrunmented with AspectC++, in this case, such that the monitor is its \textit{friend} class, so the monitor can access the private attributes of \textit{Consumer}. The monitor defines a callback function being called immediately at the end of simulation-cycle notification, and a pointer pointing to an instance of \textit{Consumer}. The execution of the callback function consists of getting the current value of the received character by the consumer, assigning this character to $c\_read$, and executing the monitor one step (i.e., creating a new state and reporting it to the Plasma plugin). In case temporal resolutions defined by using the kernel simulation phases or the event notification, the calling mechanism of the callback function is realized by modification the kernel (i.e., at the end of simulation-cycle segment code, a call to the callback function is added).

The statistical model checker is implemented as a plugin of Plasma Lab~\cite{bcl13} that establishes a communication, in which the generated monitor transmits execution traces of the MUV. In the current version, the communication is done via the standard input and output. When a new state is requested, the monitor reports the current state (the values of variables in $V_{obs}$) to the plugin. The length of traces depends on the satisfaction of the formula to be verified, which is finite because the temporal operators are bounded. Similarly, the required number of execution traces depends on the hypothesis testing algorithms in use (e.g., sequential hypothesis testing or 2-sided Chernoff bound). The full implementation can be downloaded on the website of Plasma Lab\footnote{\url{https://project.inria.fr/plasma-lab/download/plugins/}}.
\subsection{Running Verification}
Running the verification tool consists of two steps as follows. First, users define a set of observed variables and a temporal resolution in a configuration file, as well as other necessary information as an input for MAG to generate a monitor and an aspect-advices file. AspectC++ then is used to instrument automatically the model. The instrumented model and the generated monitor are compiled and linked together with the SystemC kernel into an executable model in order to make a set of execution traces. 

Refering to the running example, users will define the set of observed variables $V_{obs} = \{c\_read, n_{elements}, end\_sc\}$, where \textit{c\_read} is the character read in the FIFO, \textit{$n_{elements}$} is the number of characters in the FIFO buffer, and \textit{end\_sc} is \textit{true} whenever the kernel phase is at the end of the simulation-cycle notification phase. The temporal resolution will be defined as $\mathcal{T}_r =$ $\{end\_sc\}$, meaning that a new state in execution traces is produced whenever the simulation kernel is at the end of simulation-cycle notification phase or every one nanosecond in the example since the time unit is one nanosecond. The full configuration is given as follows.
\begin{lstlisting}[caption=The configuration file for MAG,label=lst:configurationfile,mathescape]
# Where to output the monitor
output_file   ./monitor.cpp

# The (class) name of the generated monitors
mon_name   monitor

# Plasma project file
plasma_file   /PLASMA_Lab-1.3.0/fifo/fifo.plasma

# Plasma project name
plasma_project_name   fifo

# Plasma model name
plasma_model_name   fifo_model

# Instrumented executable SystemC model
plasma_model_content   /PLASMA_Lab-1.3.0/fifo/fifo

# Set to write traces to a file
write_to_file   false

# Declare monitors as friend to adder class
usertype   Consumer
usertype   Producer

# Example of how to declare type of non-native variables
type Consumer   *pnt_con
type Producer   *pnt_pro

# Module attributes
attribute pnt_con->c_int   c_read
attribute pnt_pro->c_int   c_write

# Attribute type
att_type int   c_read
att_type int   c_write
att_type int   n_elements

# Time resolution
time_resolution   MON_TIMED_NOTIFY_PHASE_END

# Properties
formula G<=#10000((c_read = 38) => (F<=#15(c_read = 64)))

# Includes the files
include consumer.h
include producer.h
\end{lstlisting}

In the second step, the plugin is used to verify the properties of interest. The satisfaction checking of the properties is brought out based on the set of execution traces by executing the instrumented SystemC model and can be done by several hypothesis testing algorithms provided by Plasma Lab.

\newpage
\section{Experimental Evaluation}
We report the experimental results for the running example and also demonstrate the use of our verification tool to analyze the dependability of a large embedded control system. The number of components in this system makes numerical approaches such as PMC unfeasible. In both case studies, we used the 2-sided Chernoff bound algorithm with the absolute error $\epsilon = 0.02$ and the confidence $\alpha = 0.98$. The experiments were run on machine with Intel Core i7 2.67 GHz processor and 4GB RAM under the Linux OS with SystemC 2.3.0, in which the checking of the properties in the running example took from less than one minute to several minutes. The analysis of the embedded and control system case study takes almost $2$ hours, in which $90$ properties were verified.
\subsection{Producer and Consumer}
Let us go back to the running example in Section \ref{sec:example}, recall that we want to compute the probability that the following property $\varphi$ satisfies every $1$ nanosecond, with the absolute error $0.02$ and the level of confidence $0.98$. In this verification, both the FIFO buffer size and message size are $10$ characters including the starting and ending delimiters, and the production and consumption rates are $1$ nanosecond.
\begin{displaymath}
\varphi = G_{\leq T}((c\_read = \; '\&') \rightarrow F_{\leq T_1}(c\_read = \; '@'))
\end{displaymath}

First, we check this property with the various values of $p_1$ and $p_2$. The results are given in Table~\ref{tab:fifolatency} with $T = 5000$ and $T_1 = 25$ nanoseconds. It is trivial that the probability that the message latency is smaller than $T_1$ time increases when $p_1$ and $p_2$ increase. That means that, in general, the latency is shorter when the either the probability that the producer successfully writes to the FIFO increases, or the probability that the consumer successfully reads from the FIFO increases. 
\begin{table}[ht]
\centering
\begin{tabular}{c|ccc}
$p_1$\textbackslash $p_2$ & $0.3$ & $0.6$ & $0.9$\\
\hline
$0.6$ & $0$ & $0.0194$ & $0.0720$\\
$0.9$ & $0$ & $0.0835$ & $1$\\
\end{tabular}
\caption{The probability that the message latency is smaller than $25$ in the first $5000$ nanoseconds of operation}
\label{tab:fifolatency}
\end{table}
\begin{figure}[ht]
\begin{center}
\includegraphics[width=0.90\textwidth]{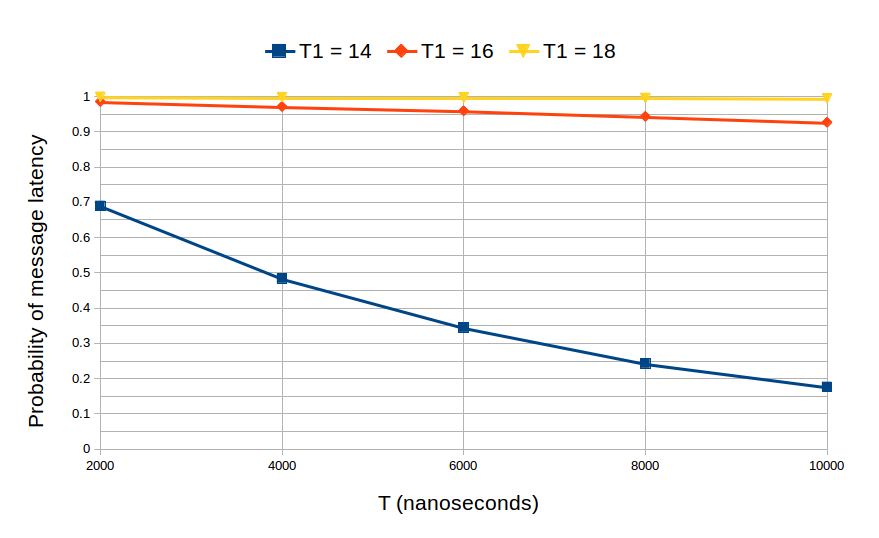}
\caption{The probability that the message latency is smaller than $T_1$ in the first $T$ nanoseconds of operation}
\label{fig:fifolatency}
\end{center}
\end{figure}

Second, we compute the probability that a message is sent completely (or the message latency) from the producer to the consumer within $T_1$ time over a period of $T$ time of operation, in which the probabilities $p_1$ and $p_2$ are fixed at $0.9$. Fig.~\ref{fig:fifolatency} shows this probability with different values of $T_1$ over $T = 10000$ nanoseconds. It is observed that the message latency is almost smaller than $18$ nanoseconds.
\subsection{An Embedded Control System}
\label{subsec:ecs}
This case study is closely based on the one presented in \cite{mct94,knp07} but contains much more components. The system consists of an input processor ($I$) connected to $50$ groups of $3$ sensors, an output processor ($O$), connected to $30$ groups of $2$ actuators, and a main processor ($M$), that communicates with $I$ and $O$ through a bus. At every cycle, 1 minute, the main processor polls data from the input processor that reads and processes data from the sensor groups. Based on this data, the main processor constructs commands to be passed to the output processor for controlling the actuator groups. 

The reliability of the system is affected by the failures of the sensors, actuators, and processors. The probability of bus failure is negligible, hence we do not consider it. The sensors and actuators are used in $37-\text{of}-50$ and $27-\text{of}-30$ modular redundancies, respectively. That means if at least $37$ sensor groups are functional (a sensor group is functional if at least $2$ of the $3$ sensors are functional), the system obtains enough information to function properly. Otherwise, the main processor is reported to shut the system down. In the same way, the system requires at least $27$ functional actuator groups to function properly (a actuator group is functional if at least $1$ of the $2$ actuators is functional). Transient and permanent faults can occur in processors $I$ or $O$ and prevent the main processor($M$) to read data from $I$ or send commands to $O$. In that case, $M$ skips the current cycle. If the number of continuously skipped cycles exceeds the limit $K$, the processor $M$ shuts the system down. When a transient fault occurs in a processor, rebooting the processor repairs the fault. Lastly, if the main processor fails, the system is automatically shut down. The mean times to failure for the sensors, the actuators, and the processors are 1 month, 2 months and 1 year, respectively. The mean time to transient failure is 1 day and I/O processors take 30 seconds, 1 time unit, to reboot.

The reliability of the system is modeled as a CTMC~\cite{mge82,tks82,gaa87} that is realized in SystemC, in which a sensor group has $4$ states ($0, 1, 2, 3$, the number of working sensors), $3$ states ($0, 1, 2$, the number of working actuators) for an actuator group, $2$ states for the main processor ($0$: failure, $1$: functional), and $3$ states for I/O processors ($0$: failure, $1$: transient failure, $2$: functional). A state of the CTMC is represented as a tuple of the component's states, and the mean times to failure define the delay before which a transition between states is enabled. The delay is sampled from a negative exponential distribution with parameter equal to the corresponding mean time to failure. Hence, the model has about $2^{155}$ states comparing to the model in \cite{knp07} with about $2^{10}$ states, that makes the PMC technique is unfeasible. That means the state explosion likely occurs, even with some abstraction, i.e., symbolic model checking is applied. The full implementation of the SystemC code of this case study can be obtained at the website of our tool\footnote{\url{https://project.inria.fr/plasma-lab/embedded-control-system/}}.

We define four types of failures: $failure_1$ is the failure of the sensors, $failure_2$ is the failure of the actuators, $failure_3$ is the failure of the I/O processors and $failure_4$ is the failure of the main processor. For example, $failure_1$ is defined by $number\_sensors < 37) \wedge (proci\_status = 2)$. It specifies that the number of working sensor groups has decreased below $37$ and the input processor is functional, so that it can report the failure to the main processor. We define $failure_2$, $failure_3$, and $failure_4$ in a similar way. 

\begin{figure}[ht]
\begin{center}
\includegraphics[width=0.90\textwidth]{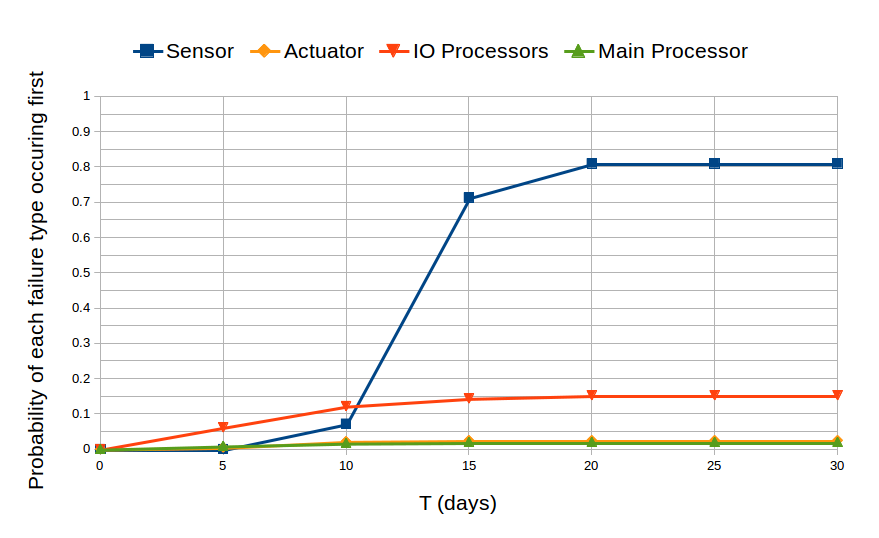}
\caption{The probability that each of the 4 failure types is the cause of system shutdown in the first $T$ time of operation}
\label{fig:failurefirst}
\end{center}
\end{figure}

In our analysis which is based on the one in \cite{knp07} with $K = 4$, in which the properties are checked every $1$ time unit. First, we try to determine which kind of component is more likely to cause the failure of the system, meaning that we determine the probability that a failure related to a given component occurs before any other failures. The atomic proposition $shutdown = \bigvee_{i=1}^4failure_i$ indicates that the system has shut down because one of the failures has occurred, and the BLTL formula $\neg shutdown \; U_{\leq T} \; failure_i$ states that the failure $i$ occurs within $T$ time units and no other failures have occurred before the failure $i$ occurs. Fig.~\ref{fig:failurefirst} shows the probability that each kind of failure occurs first over a period of $30$ days of operation. It is obvious that the sensors are likelier to cause a system shutdown. At $T=20$ days, it seems that we reached a stationary distribution indicating for each kind of component the probability that it is responsible for the failure of the system.

\begin{figure}[ht]
\begin{center}
\includegraphics[width=0.90\textwidth]{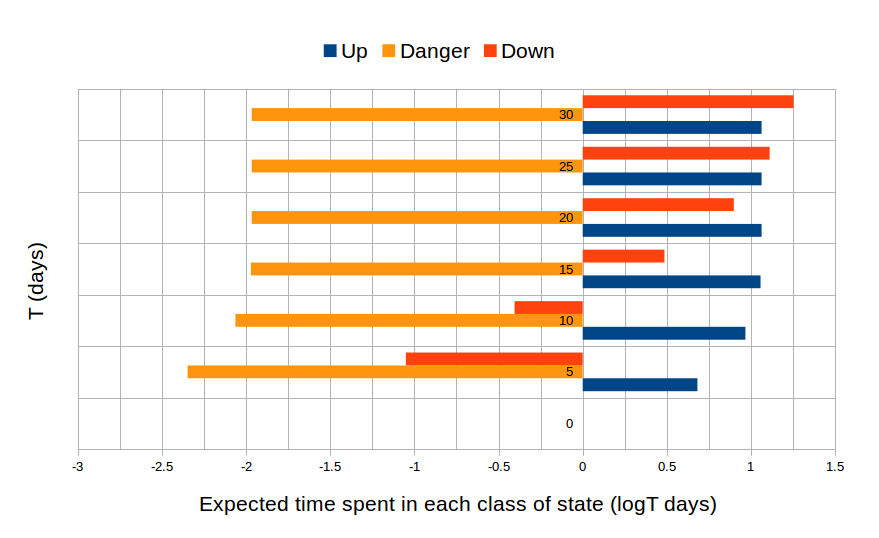}
\caption{The expected amount of time spent in each of the states: ``up'', ``danger'' and ``shutdown''}
\label{fig:timespent}
\end{center}
\end{figure}

For the second part of our analysis, we divide the states of system into three classes: ``up'', where every component is functional, ``danger'', where a failure has occurred but the system has not yet shut down (e.g., the I/O processors have just had a transient failure but they have rebooted in time), and ``shutdown'', where the system has shut down \cite{knp07}. We aim to compute the expected time spent in each class of states by the system over a period of $T$ time units. To this end, we add in the model, for each class of state $c$, a random variable $reward\_c$ that measures the time spent in the class $c$. 
In our tool, the formula $X_{\leq T} \; reward\_c$ returns the mean value of $reward\_c$ after $T$ time of execution. The results are plotted in Fig.~\ref{fig:timespent}. From $T=20$ days, it seems that the amounts of time spent in the ``up'' and ``danger'' states are converged at $10^{1.063} = 11.57$ days and $10^{-1.967} = 0.01$ days, respectively. Due to the lack of space, we present the other parts of the analysis in Appendix B.
We also study the probability that each of the four types of failure eventually occurs in the first $T$ time of operation. This is done using the BLTL formula $F_{\leq T} \; (failure_i)$.
\begin{figure}[ht]
\begin{center}
\includegraphics[width=0.90\textwidth]{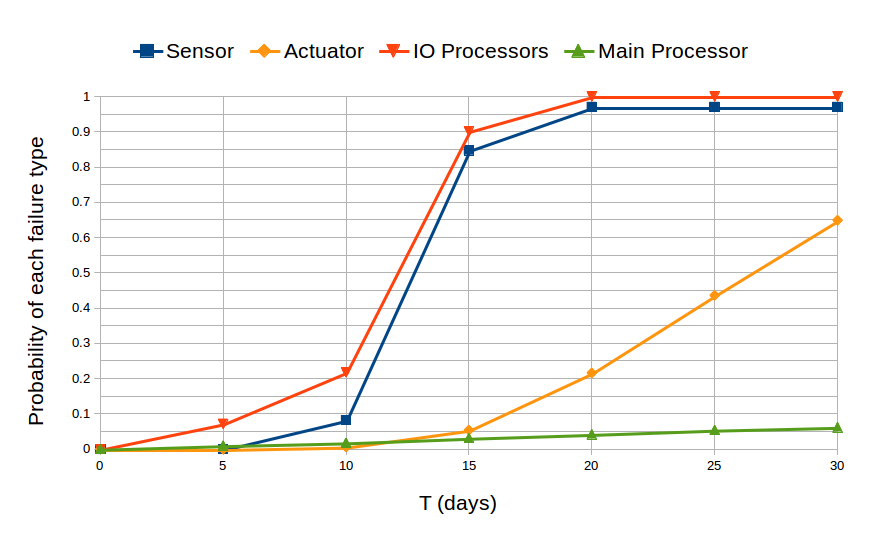}
\caption{The probability that each of the 4 failure types in the first $T$ time of operation}
\label{fig:failure}
\end{center}
\end{figure}
\begin{figure}[ht]
\centering
\includegraphics[width=0.90\textwidth]{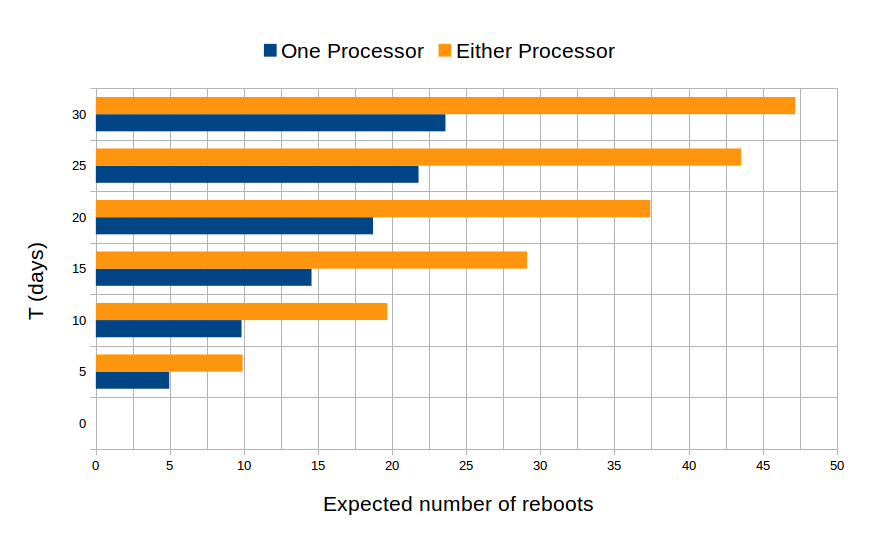}
\caption{Expected number of reboots that occur in the first $T$ time of operation}
\label{fig:numberofreboots}
\end{figure}
\begin{figure}[ht]
\centering
\includegraphics[width=0.90\textwidth]{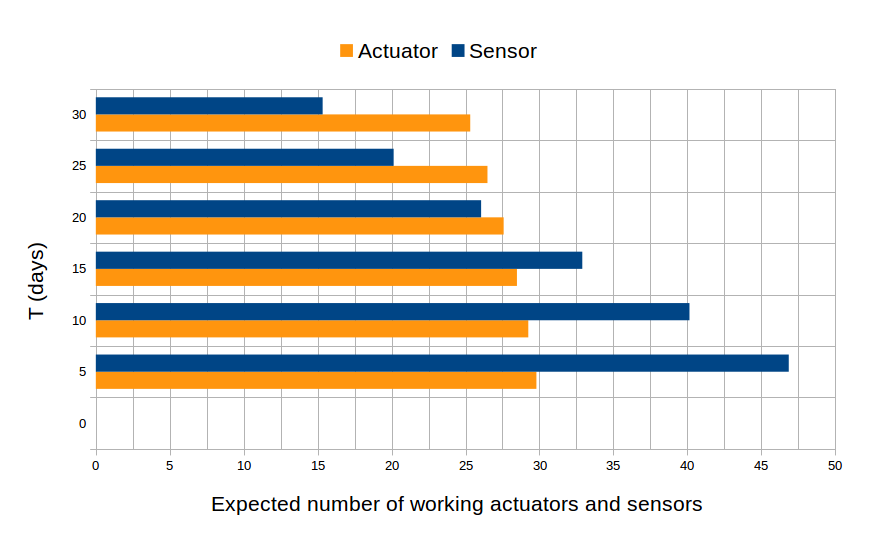}
\caption{Expected number of functional sensor and actuator groups in the first $T$ time of operation}
\label{fig:numberofworkinggroups}
\end{figure}
\figref{fig:failure} plots these probabilities over the first $30$ days of operation. We observe that the probabilities that the sensors and I/O processors eventually fail are more than the others do. In the long run, they are almost the same and approximate to $1$, meaning that the sensors and I/O processors will eventually fail with probability $1$. The main processor has the smallest probability to eventually fail.

Finally, we approximate the number of reboots of the I/O processors, and the number sensor groups, actuator groups that are functional over time by computing the expected values of random variables that count the number of reboots, functional sensor and actuator groups. The results are plotted in Fig. \ref{fig:numberofreboots} and Fig. \ref{fig:numberofworkinggroups}. It is obvious that the number of reboots of both processors doubles the number of reboots of each processor since they have the same behavior.

%% file: conclusion.tex
\section{Conclusions}
There has been a lot of work on the formalization of SystemC~\cite{hfg08,mmh08}. The goal is to extract a formal model from a SystemC program, so that tools like model-checkers can be applied. However, all these formalizations consider semantics of SystemC and its simulator in some form of \textit{global model}, and they also suffer from the state space explosion when dealing with industrial and large systems.

Tabakov et al.~\cite{tva10} proposed a framework for monitoring temporal SystemC properties. This framework allows users express the verifying properties by fully exposing the semantics of the simulator as well as the user-code. They extend LTL by providing some extra primitives for stating the atomic propositions and let users define a much finer temporal resolution. Their implementation consists of a modified simulation kernel, and a tool to automatically generate the \textit{monitors} and aspect advices for applying \textit{Aspect Oriented Programming} (AOP) \cite{gss01} to instrument SystemC programs automatically.

This paper presents the first attempt to verify non-trivial temporal properties of SystemC model with statistical model checking techniques. The framework contains two main components: a \textit{generator} that automatically generates a monitor and instruments the MUV based on the properties to be verified, and a \textit{statistical model checker} implementing a set of hypothesis testing algorithms. In comparison to the probabilistic model checking, our approach allows users to handle large industrial systems, expose a rich set of user-code primitives in form of atomic propositions in BLTL, and work directly with SystemC models. For instance, our verification framework is used to analyze the dependability of large industrial computer-based control systems as shown in the case study.

Currently, we consider an external library as a ``black box'', meaning that we do not consider the states of external libraries. Thus, arguments passed to a function in an external library cannot be monitored. For future work, we would like to allow users to monitor the states of the external libraries. We also plan to apply statistical model checking to verify temporal properties of SystemC-AMS (Analog/Mixed-Signal).